\documentclass[12pt]{article} \topmargin= 0.3cm \textwidth 16.5cm
=0cm \headsep =0cm
\usepackage {latexsym}
\usepackage {graphicx}

\begin{document}
\title{Brane world scenario in the presence of a non-minimally coupled bulk scalar field}
\author{K. Farakos \footnote{kfarakos@central.ntua.gr} and P. Pasipoularides  \footnote{paul@central.ntua.gr} \\
       Department of Physics, National Technical University of
       Athens \\ Zografou Campus, 157 80 Athens, Greece\\ (talk given by P. Pasipoularides in Nafplio NEB XII 2006)}
\date{ }
       \maketitle

\begin{abstract}
We present our recent work on brane world models with a
non-minimally coupled scalar field. In \cite{FP} we examined the
stability of these models against scalar field perturbations and we
discussed possible physical implications, while in \cite{KF} we
developed a numerical approach for the solution of the Einstein
equations with the non-minimally coupled scalar field.
\end{abstract}
%%%%%%%%%%%%%%%%%%%%%%%%%%%%%%%%%%%%%%%%%%%%%%%%%%%%%%%
\section{Introduction}

A possible way to extend four-dimensional physics is to consider
extra dimensions. However, our world appears to be four dimensional,
thus extra dimensions should be hidden in low energies. Two
mechanisms for this are available: a) The Kaluza-Klein Scenario
(small compact extra dimensions with compactification scale
$M>1TeV$) and b) The brane world scenario.

According to the brane world scenario, ordinary matter is assumed to
be trapped in a 3D submanifold (brane world) that is embedded in a
multi-dimensional manifold (bulk). Contrary to ordinary matter,
gravitons are allowed to propagate in the bulk. The brane world
scenario has the advantage that it predicts new phenomenology even
at the TeV scale. Furthermore, it puts on a new basis fundamental
problems such as the hierarchy and the cosmological constant
problem.

For the localization of ordinary matter on the brane we can mention
two mechanisms. According to the first one, fermions are trapped on
the brane due to the existence of a topological defect toward the
extra dimensions \cite{Rub} . However, there is a more powerful
mechanism \cite{Shif}, for the localization of extended structures
of particles which includes: Gauge fields, fermions and bosons with
Gauge charge. This mechanism is based on a specific phase structure
which is known as the layer phase (Higgs phase along the brane
combined with a confinement phase along the extra dimensions), see
details in Refs. \cite{lat6:KA,lat}.

The standard brane world scenario with gravity and flat compact
extra dimensions is the ADD model \cite{ADD} (Arkani-Hammed,
Dimopoulos and D'vali). In the case of warped extra dimensions we
have the first and second Randall-Sundrum models  \cite{Ran}.

According to the second Randall-Sundrum model (RS2-model), we have a
single brane with a positive energy density (the tension $\sigma$),
whereas the bulk has a negative five-dimensional cosmological
constant $\Lambda$. The corresponding Einstein equations have a
solution only if a fine-tuning condition is satisfied
($\Lambda=-\frac{\sigma^{2}}{6}$), in units where $8 \pi G_5=1$. An
extension of the RS2-model with a second negative tension $-\sigma$
brane, is the RS1-model \cite{Ran}. In this case we have an
orbifolded extra dimension of radius $r_c$. The two branes are
fitted to the fixed points of the orbifold, $z=0$ and $z_c=\pi r_c$
with tensions $\sigma$ and $-\sigma$ correspondingly. The particles
of the standard model are assumed to be trapped on the negative
tension brane, which is called visible, while the positive tension
brane is called hidden.

There are numerous generalizations of the standard Randall-Sundrum
scenario, like models with more than five dimensions, models with
topological defects toward the extra dimensions, multibrane models
and models with higher order curvature corrections (i.e.
Gauss-Bonnet gravity). See for example \cite{Rub,Lon,AH} and
references therein.

In our recent papers \cite{FP,KF} we have studied a generalization
of the RS2-model with a nonminimally coupled bulk scalar field, via
an interaction term of the form $-\frac{1}{2}\xi R \phi^2$, where
$\xi$ is a dimensionless coupling \footnote{The coupling $\xi$
possesses two characteristic values: a) the minimal coupling for
$\xi=0$ and b) the conformal coupling for $\xi_{c}=3/16$ }. In
particular, the stability properties of the model are investigated
in Ref. \cite{FP}, and a possible implication of the model to the
layer phase mechanism for the localization of ordinary matter on the
brane is discussed. Furthermore, in Ref. \cite{KF} we solve
numerically the Einstein equations with the non-minimally coupled
scalar field and the appropriate boundary conditions on the brane.
Although our numerical approach is suitable for an arbitrary form of
the potential, we have examined the simplest case of a potential
$V(\phi)=\lambda \phi^4$ for the scalar field. We showed that,
according to the value of the nonminimal coupling $\xi$, our model
possesses three classes of new static solutions (for details see
section 2.5).

It is worth to mention, that in the case of brane models with a
non-minimally coupled scalar field, K. Tamvakis and collaborators
have found analytical solutions by choosing appropriately the
potential for the scalar field \cite{Tam}.

Furthermore, for implications of brane world models with a
non-minimally coupled scalar field to the stabilization of the extra
dimension you can see \cite{GW,Cas}, while for phenomenological
implications the reader may consult \cite{Toms,Dav}.

5D models in the framework of Brans-Dicke theory have been also
studied. Particularly, in Ref. \cite{bd}, static analytical
solutions are constructed for a special class of potentials, and the
stabilization of the extra dimension is discussed.

\section{Brane world models with a nonminimally coupled scalar field}

\subsection{The action of the model}

In \cite{FP,KF} we have studied brane world models with a
nonminimally coupled bulk scalar field. The action of these models
is:
\begin{equation}
 S=\int d^{5}x \;({\cal L}_{RSF}+{\cal L}_{\phi})
\end{equation}
where $d^{5}x=d^{4}x dz$, and $z$ parameterizes the extra dimension.

The gravity part of the lagrangian, if we set $8\pi G_{5}=1$ ($G_5$
is the five-dimensional Newton constant), is given by the equation
\begin{equation}
{\cal L}_{RSF}=\sqrt{|g|}\left(
F(\phi)R-\Lambda\right)-\sigma\delta(z)\sqrt{|g^{(brane)}|}
\end{equation}
where $R$ is the five-dimensional Ricci scalar, $g$ is the
determinant of the five-dimensional metric tensor $g_{MN}$ ($M,
N=0,1,...,4$), and $g^{brane}$ is the determinant of the induced
metric on the brane. We adopt the mostly plus sign convention for
the metric \cite{R2}. Furthermore, $\Lambda$ is a negative five
dimensional cosmological constant, and $\sigma$ is the brane
tension.

The factor
\begin{equation}
F(\phi)=\frac{1}{2}(1-\xi \phi^2 )
\end{equation}
corresponds to a nonminimally coupled scalar field with an
interaction term of the form  ${\cal L}_{int}=-\frac{1}{2}\xi R\;
\phi^{2}$, and $\xi$ is a dimensionless coupling.  Note that if we
set $\xi=0$ the model is reduced to the RS2-model with a minimally
coupled scalar field.

The scalar field part of the lagrangian is
\begin{eqnarray}
{\cal L}_{\phi}&=&\sqrt{|g|}\left(-\frac{1}{2}g^{M N} \nabla_{M}\phi
\nabla_{N}\phi-V(\phi)\right)
\end{eqnarray}
where the potential is assumed to be of the standard form
$V(\phi)=\lambda \phi^{4}$.

\subsection{An obvious solution of the model in the case of the fine tuning $\Lambda=\frac{-\sigma^2}{6}$}

If the fine tuning $\Lambda=\frac{-\sigma^2}{6}$ is satisfied, the
Einstein equations have a \textit{static solution} of the form
\begin{equation}
ds^{2}=a(z)^2(-dx_{0}^{2}+dx_{1}^{2}+dx_{2}^{2}+dx_{3}^{2})+dz^{2}
\end{equation}
where $a(z)=e^{-k|z|}$ is the warp factor and
$k=\sqrt{\frac{-\Lambda}{6}}$.

We recognize the well known solution of the RS2-model. However, in
the case we examine there is a difference, as for appropriate values
of the dimensionless coupling $\xi$ this solution is unstable
against perturbations of the scalar field \cite{FP}, see for details
the next subsection.

\subsection{Spectrum of the scalar field in the presence of the RS2-metric}

In Refs. \cite{FP,KF} we studied the spectrum of scalar field in the
background of the RS2-vacuum. In this section we give the results of
our study.

If we set
\begin{equation}
w=sgn(z)\frac{(e^{k|z|}-1)}{k}
\end{equation}
the RS2-metric of Eq. (5) can be put into the manifestly conformal
to the five-dimensional Minkowski space form
\begin{equation} ds^{2}=\alpha(w)^2(-dx_{0}^{2}+dx_{1}^{2}+dx_{2}^{2}+dx_{3}^{2}+dw^{2})
\end{equation}
where
\begin{equation} \alpha(w)=
\frac{1}{k|w|+1}
\end{equation}

If we consider a perturbation $\hat{\phi}$ around the scalar field
vacuum ($\phi=0$) we obtained the corresponding linearized equation:
\begin{equation}
\frac{1}{\sqrt{|g|}}\;
\partial_{M}\left[\sqrt{|g|}g^{MN}\partial_{N}\hat{\phi}(x,w)\right]+\xi
R(w) \;\hat{\phi}(x,w)=0
\end{equation}

We can set
\begin{equation}
\hat{\phi}(x,w)=e^{ipx} \frac{\psi(w)}{\alpha^{3/2}(w)}
\end{equation}
where $\alpha(w)=1/(k|w|+1)$, and $m^{2}=p_{\mu}p^{\mu}$ is the
effective four dimensional mass.

The function $\psi(w)$ satisfies the Schrondiger-like equation
\begin{equation}
-\psi''(w)+\left[V(w)-m^{2}\right]\psi(w)=0
\end{equation}
where the potential $V(w)$ is equal to
\begin{equation}
V(w)=\frac{(\alpha^{3/2}(w))''}{\alpha^{3/2}(w)}+\xi\alpha^{2}(w)R(w)
\end{equation}
From Eqs. (8) and (12) we get
\begin{equation}
V(w)= -16 k(\xi-\xi_c)\left(-\delta(w)+\frac{5
k}{4(k|w|+1)^{2}}\right)
\end{equation}
where $\xi_c=3/16$ is the five dimensional conformal coupling.

Note that the coefficient in front of the potential change sign when
$\xi$ crosses the five dimensional conformal coupling. This result
implies that the potential has two characteristic forms, as we see
in the left-hand panel ($\xi<\xi_c$) and the right-hand panel
($\xi>\xi_c$) of Fig. 10 in \cite{KF}.

If we use the above results, Eqs. (11) and (13), we can show (for
details see Refs. \cite{FP,KF}) that:

\begin{enumerate}
    \item For $\xi<0$ there is a unique tachyon mode, localized on the brane, plus a continuous spectrum of positive energy states.
    \item For $\xi>\xi_c$ there is at least one tachyon mode or more than one, depending on the value of $\xi$, plus a continuous spectrum of positive energy
    states.
    \item For $0<\xi\leq \xi_c$ there are no tachyon modes. There is only a continuous spectrum of positive energy
    states.
    \item For $\xi=0$ there is a zero mode, plus a continuous spectrum of positive energy states.
\end{enumerate}

The above results implies an instability of the RS2-metric for
$\xi>\xi_c$ or $\xi<0$.

\subsection{Einstein equations with the non-minimally coupled scalar field}

The Einstein equations, which correspond to the action of Eq. (1)
are
\begin{eqnarray}
G_{MN}+\Lambda\; g_{MN}+\sigma
\delta(z)\frac{\sqrt{|g^{(brane)}|}}{\sqrt{|g|}}g_{\mu\nu}\delta_{M}^{\mu}\delta_{N}^{\nu}=T^{(\phi)}_{MN}
\end{eqnarray}
where the energy momentum tensor for the scalar field is
\begin{equation}
T^{(\phi)}_{MN}=\nabla_{M}\phi\nabla_{N}\phi-g_{MN}[\frac{1}{2}g^{P\Sigma}\nabla_{P}\phi\nabla_{\Sigma}\phi+V(\phi)]+2
\nabla_{M}\nabla_{N}F(\phi)-2 g_{MN}\Box F(\phi)+(1-2 F(\phi))G_{MN}
\end{equation}

The equation of motion for the scalar field is
\begin{equation}
\Box \phi+\frac{\partial F(\phi)}{\partial \phi} R-\frac{\partial
V(\phi)}{\partial \phi}=0
\end{equation}
The above equation is not independent of the Einstein equations (7),
as it is equivalent to the conservation equation $\nabla^M
T^{(\phi)}_{MN}=0$, where $T^{(\phi)}_{MN}$ is given by Eq. (15).

We are looking for static solutions of the form
\begin{equation}
ds^{2}=a^{2}(z)(-dx_{0}^{2}+dx_{1}^{2}+dx_{2}^{2}+dx_{3}^{2})+dz^{2},
\quad \phi=\phi(z)
\end{equation}

We obtain (for detail see Ref. \cite{KF}) the following second order
differential equations:
\begin{eqnarray}
3(1-\xi\phi^{2}(z)) A''(z)+(1-2 \xi)\phi'(z)^2-2 \xi \phi(z)
\phi''(z)+2 \xi A'(z)\phi(z) \phi'(z)+\sigma \delta(z)=0
\end{eqnarray}
\begin{equation}
-\phi''(z)-4 A'(z) \phi'(z)-\xi\left(8
A''(z)+20A'(z)^2\right)\phi(z)+V'(\Phi)=0
\end{equation}
and the constraint equation
\begin{eqnarray}
6(1-\xi\phi^{2}(z))
A'(z)^2+\Lambda-\frac{1}{2}\phi'(z)^2+V(\phi(z))-8 \xi A'(z)\phi(z)
\phi'(z)=0
\end{eqnarray}
Note that only two of the above three equations are independent.

As this system is complicated we will not look for analytical
solutions, but we will try to solve it numerically.

For the numerical integration of these equations it is necessary to
know the values of $A(0)$, $\phi(0)$, $A'(0)$ and $\phi'(0)$. These
values are determined by the junction conditions (see Eqs. (18),
(19) and the constraint equation.

According to the junction conditions
\begin{equation}
A'(0)=\frac{-\sigma}{(6-6 \xi \phi(0)^2+32 \xi^2 \phi(0)^2)}
\end{equation}
\begin{equation}
\phi'(0)=\frac{8 \xi \sigma \phi(0)}{(6-6 \xi \phi(0)^2+32 \xi^2
\phi(0)^2)}
\end{equation}
while $\phi(0)$ can be obtain by the following sixth order algebraic
equation:
\begin{equation}
\frac{\sigma^2}{(6-6 \xi \phi(0)^2+32 \xi^2
\phi(0)^2)}+\Lambda+V(\phi(0))=0
\end{equation}

\subsection{Numerical solutions}

For the determination of the two unknown functions $A(z)$ and
$\phi(z)$ we can solve the system of second order differential
equations (18) and (19) for $z\geq 0$ numerically. In order to
integrate it is necessary to know the values of $A(0)$, $\phi(0)$,
$A'(0)$ and $\phi'(0)$. If we assume that the warp factor is
normalized to unity on the brane (or $a(0)=1$) we find that $A(0)=0$
(note that $a(z)=e^{A(z)}$). The value of $\phi(0)$ is obtained by
solving Eq. (22), and the values of $A'(0^+)$ and $\phi'(0^+)$ can
be found from Eqs. (20) and (21). Then it is an easy task to use a
routine of Fortran or Mathematica to extract the numerical
solutions.

The model we examine has four independent parameters
$\xi,\lambda,\Lambda,\sigma$. We kept fixed the parameters
$\lambda,\Lambda,\sigma$, assuming that the fine tuning
$\Lambda=-\frac{\sigma^2}{6}$ is satisfied, and we varied the
parameter $\xi$. Depending on the value of $\xi$ we obtained three
classes of numerical solutions with different characteristics:

\begin{enumerate}
    \item For $\xi<0$ the warp factor exhibits a naked
singularity at finite proper distance $z_{s}$ in the bulk, while the
scalar field $\phi(z)$ is almost constant near the brane, and tends
to infinity as z tends to the singularity point in the bulk. See
Figs. 1,2,3 in \cite{KF}.
    \item For $\xi>\xi_{c}$ the warp factor $a(z)$ is of
the order of unity in a small region near the brane and increases
exponentially ($a(z)\rightarrow e^{k z}$ with
$k=\sqrt{\frac{-\Lambda}{6}}$), as $z\rightarrow +\infty$ , while
the scalar field $\phi(z)$ is nonzero on the brane and tends rapidly
to zero in the bulk. See Figs. 4,5,6 in \cite{KF}.
    \item For $0<\xi<\xi_{c}$ the warp factor $a(z)$ of this solution
tends rapidly to infinity (faster than case (2)), also the scalar
field $\phi(z)$ is nonzero on the brane and tends rapidly to
infinity in the bulk.  Contrary to case (2), where the space-time is
asymptotically $AdS_{5}$, in this case the scalar curvature tends to
infinity. See Fig. 7 in \cite{KF}.
\end{enumerate}

Also we investigated what happens when the fine tuning is violated,
and we found that in appropriate regions of the parameter space, the
three classes of solutions we described above are preserved.
However, there are regions of the parameters where there are no
static solutions (or Eq. (22) has no real solutions). A thorough
investigation of these regions is very extended and it is beyond the
scope of our work.

Furthermore, in \cite{KF} we have examined the stability of the
above mentioned solutions and we obtained that the first and second
classes are unstable against scalar field perturbations, while the
third class is stable.

\begin{figure}[h]
\begin{center}
\includegraphics[scale=1.2,angle=0]{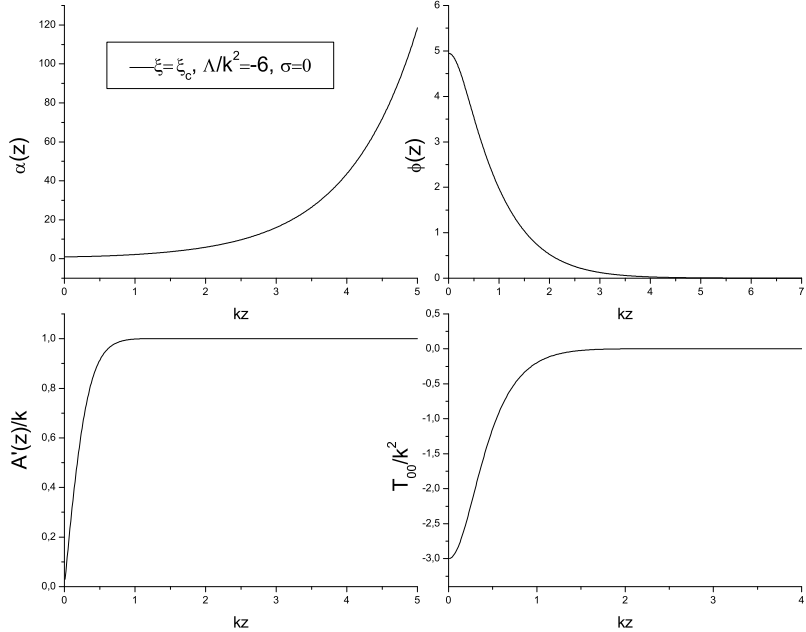}
\end{center}
\caption {$a(z)$, $\phi(z)$, $A'(z)/k$ and $T_{00}(z)/k^2$, for
$\xi=\xi_c$, as a function of kz for $\sigma=0$ and
$\lambda/k^2=0.01$, where $k=\sqrt{\frac{-\Lambda}{6}}$.}
\label{con2}
\end{figure}

Separately, we have examined the case of conformal coupling $\xi_c$,
and we saw that the corresponding solutions are the form of the
second class (see Fig. \ref{con2} for a tensionless brane). However,
in this case the solutions are found to be stable. Possible physical
implications are discussed in \cite{KF}.

\section{Acknowledgements} This work is supported by the EPEAEK programme "Pythagoras"
and co-founded by the European-Union (75$^{\circ}/_{\circ}$) and the
Hellenic state (25$^{\circ}/_{\circ}$)

\end{document}